# Positive attitudinal shifts and a narrowing gender gap: Do expertlike attitudes correlate to higher learning gains for women in the physics classroom?

Alma Robinson[,1,*] John H. Simonetti[,1] Kasey Richardson[,2] and Megan Wawro[3]

[1]*Department of Physics, Virginia Tech, Blacksburg, Virginia 24061, USA*
[2]*School of Education, Virginia Tech, Blacksburg, Virginia 24061, USA*
[3]*Department of Mathematics, Virginia Tech, Blacksburg, Virginia 24061, USA*

A large body of research shows that using interactive engagement pedagogy in the introductory physics classroom consistently results in significant student learning gains; however, with a few exceptions, those learning gains tend not to be accompanied by more expertlike attitudes and beliefs about physics and learning physics. In fact, in both traditionally taught and active learning classroom environments, students often become more novicelike in their attitudes and beliefs following a semester of instruction. Furthermore, prior to instruction, men typically score higher than women on conceptual inventories, such as the Force Concept Inventory (FCI), and more expertlike on attitudinal surveys, such as the Colorado Learning Attitudes about Science Survey (CLASS), and those gender gaps generally persist following instruction. In this paper, we analyze three years of pre-post matched data for physics majors at Virginia Tech on the FCI and the CLASS. The courses were taught using a blended pedagogical model of peer instruction, group problem solving, and direct instruction, along with an explicit focus on the importance of conceptual understanding and a growth mindset. We found that the FCI gender gap decreased, and both men and women showed positive, expertlike shifts on the CLASS. Perhaps most surprisingly, we found a meaningful correlation between a student's post-CLASS score and normalized FCI gain for women, but not for men.

DOI: 10.1103/PhysRevPhysEducRes.17.010101

## I. INTRODUCTION

A substantial body of work in physics education research has demonstrated that interactive-engagement instructional techniques lead to significantly higher student learning gains on conceptual inventories such as the Force Concept Inventory (FCI) [1] and the Force and Motion Conceptual Evaluation (FMCE) [2] as compared to traditional, lecture-based practices [3]. However, the effect that these interactive engagement techniques have on the gender gap (the differential scores between men and women on these conceptual inventories) and student attitudes and beliefs about physics and the learning of physics has been unclear.

Madsen, McKagan, and Sayre reviewed the literature on the disparity of scores between men and women on the FCI and FMCE and found that, as a weighted average, men scored higher than women by 13% on the pretest and 12% on the post-test, and also achieved a 6% higher normalized gain [4]. Although there is some evidence that interactive engagement courses can help reduce the gender gap, other studies have shown that interactive engagement alone is insufficient [5].

A study by Lorenzo, Crouch, and Mazur [6] showed a significant narrowing of the pre to post FCI and FMCE gender gap in physics courses at Harvard University that used an array of interactive engagement strategies including Peer Instruction [7], Tutorials in Introductory Physics [8], and cooperative problem solving. However, when Pollock, Finkelstein, and Kost tried to replicate the Harvard findings at the University of Colorado by using similar pedagogical techniques, they found that although interactive engagement techniques resulted in higher normalized learning gains, the gender gap was not reduced [9]. Beichner *et al.* report that women have higher success rates in SCALE-UP classrooms than in traditional physics courses [10]. It seems that although interactive engagement courses can show encouraging steps towards eliminating the gender gap, there may be other factors at play.

There is also a large amount of research documenting how physics students' beliefs and attitudes about physics and learning physics change after taking a physics course. Two commonly used surveys that measure these beliefs are the Maryland Physics Expectations Survey (MPEX) [11] and the Colorado Learning Attitudes about Science Survey (CLASS) [12]. The CLASS has students respond to 42 statements on a Likert scale (strongly agree to strongly disagree) about aspects such as the conceptual and mathematical coherence of physics, the relevance of physics to the real world, and the effort required to understand physics and solve physics problems. A student's "percent favorable response" or "percent expertlike response" is determined by the percentage of their answers that agree with the answers given by experts (physics faculty). We would hope that our students' views of physics would become more expertlike after receiving a semester of physics instruction, but typically students shift to become more novicelike [11–14]. One might reasonably expect that interactive engagement courses would be able to defy this trend, but the courses or curricula that report positive, or expertlike, shifts [15–21] are the exception, rather than the rule. Furthermore, the CLASS authors showed that men have more expertlike pretest attitudes than women do [12] and Kost, Pollock, and Finkelstein found that although both men and women experienced negative CLASS shifts, the negative shifts for women were larger [14].

[*]alma.robinson@vt.edu

Despite these trends, however, significant positive student attitudinal gains have been documented by courses that use a variety of pedagogical strategies including Physics and Everyday Thinking [15], Modeling Instruction [16,18], Physics by Inquiry [17], and Peer Instruction [20].

Interestingly, one study in the United Kingdom by Bates, Galloway, Loptson, and Slaughter found that physics majors' attitudes, as measured by the CLASS, remained relatively unchanged during their undergraduate years [22].

In this paper, we build on this body of work [23] and describe how the blended pedagogical model that is used to teach introductory physics to physics majors at Virginia Tech, which is supplemented with a 1-credit First-Year Experience course for physics majors, has resulted in significant learning gains and a narrowing of the gender gap on the FCI, as well as expertlike attitudinal gains for both men and women on the CLASS.

## II. THE FIRST-YEAR EXPERIENCE FOR PHYSICS MAJORS

Virginia Tech is a large, public, land-grant, research university with a strong engineering program with about 2000 students taking the calculus-based introductory physics classes each semester. To help ensure that physics majors can form a cohort and be known within the physics department, they are enrolled in physics major-only sections of introductory physics and a First-Year Experience course for physics majors, both of which are taught as a two-semester sequence.

The data presented in this paper are drawn from the physics-major sections of the introductory mechanics classes from the fall semesters of 2015, 2016, and 2017. Each year, the physics majors were divided into two sections, taught by the first two authors of this paper. Their class size ranged from 34 to 71. Both instructors have extensive teaching experience.

The classes were taught in a SCALE-UP [10] classroom, where students sat at round tables of nine, but they did not strictly follow the standard SCALE-UP pedagogy. Instead, each instructor used a blended model of direct instruction, peer instruction, and group problem-solving pedagogy. Before the start of each week, the students were assigned to read the relevant sections of their textbook for the upcoming week's material and complete an online quiz. During class, the instructor interweaved minilectures, which identified essential concepts and their connection to past material, with conceptual questions that the students answered using a classroom response device (clicker). The students were asked to think about the question for a few moments on their own, then discuss their answers and reasoning with their classmates while the instructor, a graduate teaching assistant (GTA), and an undergraduate learning assistant (LA) circulated the room to help facilitate these discussions. Depending on the difficulty of the question, the instructor would sometimes probe the entire class with leading questions and/or engage the students in a classwide discussion during the process. In addition to conceptual questions, the students were also asked to solve problems, using their group members and the teaching team as resources. To ensure that the students had sufficient scaffolding to tackle these problems, the instructor modeled example problems throughout the semester, explicitly discussing the physics problem-solving process, including how to engage with mathematical sense making of equations and answers by using intuition, dimensional analysis, and checking symbolic answers with limiting cases.

Each LA was an undergraduate student who had successfully taken the course in a past semester and had also been trained (or was being trained) in physics-specific pedagogy through the department's Physics Teaching and Learning course. The GTA was either a physics graduate student or an education graduate student pursuing licensure to teach physics in secondary schools. Although not a requirement, the GTA was often a student who had chosen to take the Physics Teaching and Learning course as well. When possible, the instructors also tried to ensure that at least one member of each teaching team—the instructor, the GTA, or the LA—was a woman, and one member was a man. Of the six course sections analyzed in this study, however, one section had only men on the teaching team and one section had only women on the teaching team.

The instructors placed a large emphasis on being explicit about the value of a growth mindset [24] and the course's pedagogical choices. The teachers spend most of the first day explaining why the course focuses on conceptual understanding and active student engagement, what it means to have a growth mindset, and how those are connected to the learning process. A significant portion of each class meeting was devoted to having students work with each other to describe and/or make predictions about a physical situation using qualitative reasoning, and the benefits of these teaching approaches were identified throughout the semester.

During small group and class discussions, the instructors encouraged students to share their intuitions and reasoning with each other, and helped them make connections between their own thinking, their classmates' ideas, previous material, and the question or problem at hand. Because there are often a variety of ways to think about a physical situation, these discussions could be incredibly rich, and would sometimes involve multiple students appealing to different reasonings, both conceptual and mathematical.

Conceptual presentations of new material preceded in-depth quantitative problem solving, and the instructors would often ask for explicit mathematical sense making of equations and symbolic answers. When learning a new equation, for example, the students were sometimes asked to first guess at what it would be by appealing to their intuition. What physical quantities did they think were relevant? How were those quantities related? Directly proportional? Indirectly proportional? Something else? What units would result? Through questions, a dialogue, or a quick demonstration, the instructor attempted to help the students make sense of the mathematical representation. Similarly, after the students or the instructor reached a symbolic answer to a problem, the students were sometimes asked to discuss with their neighbors why, conceptually, the symbolic solution made sense.

The instructors also explicitly addressed the rationale, purpose, and limitations of simplifying assumptions. Without these discussions, it is not unreasonable to think that a student might leave a physics course believing that physics is not useful in the real world, where, for example, air resistance is not negligible.

Below is an example of a conceptual question that the instructors gave their introductory physics students based on a car accident that one of the instructors was involved in:

*I was on my way home from a birthday party when I drove up to a red light and stopped behind two other cars that were already waiting. A drunk driver came from behind me and slammed into the rear of my car, causing it to jolt forward into the car in front of me, which in turn, caused that car to run into the car in front of it. After realizing that I was not injured, I noticed that my glasses had flown off of my face. I found my glasses, put them on, and went outside to make sure everyone was okay. Once I confirmed that everyone was safe, I sat back down in the driver's seat and saw that my aftermarket radio had slipped out of its dashboard casing. Immediately, I thought, "Wow, that is some serious physics." When did my glasses fly off, and when did the radio slip out? Explain your reasoning.*

In addition to the three 50-min class meetings each week, the students also attended a traditional 2-h lab session and had an optional 50-min recitation per week. During recitation, the students worked on a problem set assignment, for which they produced handwritten solutions, with assistance from their peers, the GTA, and the LA. In addition to the weekly handwritten problem sets, the students also solved weekly problems using an online homework system. All students, regardless of recitation attendance, were required to complete both problem sets. Homework assignments and exams assessed both the students' conceptual understanding and their quantitative problem solving ability.

Almost all of the students in these two sections of introductory physics for physics majors were also co-enrolled in the department's 1-credit First-Year Experience course, Seminar for Physics Majors: Thinking Like a Physicist. This course met once per week and combined students from both of the aforementioned sections of introductory physics as well as first-year transfer students. The two instructors who taught the introductory physics courses team taught the seminar course using student-centered pedagogy with help from a GTA and LA.

The fall semester of the seminar course was mostly devoted to problem-solving skills, ranging from students learning how to make estimations and assumptions by solving Fermi problems, such as "*How much square footage is needed to support parking for a sold-out Virginia Tech football game*?," to approaching traditional end-of-chapter physics problems that would be found in an introductory textbook by implementing standard problem-solving techniques. The semester culminated with the students tackling an open-ended problem addressing the global energy crisis.

### III. METHODS

The participants in the study were students in the physics major sections of introductory physics during the fall semesters of 2015, 2016, and 2017. Because of the small number of women in each section, we combined the students from both classes and the three semesters into one dataset, but we have outlined each class in Table I.

TABLE I. Demographics of the student population in the physics majors' introductory physics courses ($N = 283$, 60 women, 223 men).

| Instructor | Semester | Number of students (Women, Men) |
|---|---|---|
| A | Fall 2015 | 71 (12, 59) |
| B | Fall 2015 | 64 (10, 54) |
| A | Fall 2016 | 41 (10, 31) |
| B | Fall 2016 | 39 (12, 27) |
| A | Fall 2017 | 34 (8, 26) |
| B | Fall 2017 | 34 (8, 26) |

TABLE II. Force Concept Inventory data for men and women. Pre- and postscores and normalized gains are reported ± the standard error of the mean. The 95% confidence interval (C.I.) on the effect is shown in the effect size.

| | N | Pre | Post | Normalized gain | $p$ value | Effect size |
|---|---|---|---|---|---|---|
| Men | 168 | 65.3 ± 1.5 | 84.7 ± 1.0 | 0.579 ± 0.021 | <0.0001 | 1.42 (1.18, 1.66) |
| Women | 49 | 48.6 ± 2.5 | 75.9 ± 1.9 | 0.536 ± 0.027 | <0.0001 | 2.00 (1.51, 2.50) |

University demographic data did not include gender identity, so the instructors' perceptions of students' expressed gender identities were recorded.

The students were given the FCI, a 30 question assessment that measures a student's conceptual understanding of Newtonian mechanics, as a pretest (during the first week of the semester) and as a post-test (during the last week of the semester). For our analysis, we included only matched students: those who took the FCI both pre- and postinstruction (217 students total: 168 men, 49 women). The scores are reported as the student's percentage of correct answers.

The normalized gain of each individual student, $g_I$, was calculated using

$$g_I = \frac{\text{post-test} - \text{pretest}}{100 - \text{pretest}} \quad (1)$$

and the overall normalized gain was found by averaging the individual gains. This calculation, although common in the literature [25], differs from Hake's definition of normalized gain, $\langle g \rangle$, which is calculated by calculating the gain of the class averages [3]. For classes sizes larger than 20, Hake found that these two types of averages are generally within 5% of one another [3].

The effect size was calculated using Cohen's $d$ [26], a method preferred by Nissen et al. [25] over normalized gain for measuring student learning gains, where

$$d = \frac{\text{post-test} - \text{pretest}}{\sigma_{\text{pooled}}}, \quad (2)$$

where $\sigma_{\text{pooled}}$ is the pooled standard deviation, and the angular brackets denote a mean. To report the student learning gains on the FCI in our study, we use both the effect size (Cohen's $d$) and the average of the individual normalized gains. Cohen's $d$ provides a measure of the overall learning gains for the group, but an individual student's FCI normalized gain was essential for analyzing correlations between an individual student's conceptual gains and CLASS scores.

To measure students' attitudes and beliefs about physics, the students were given the CLASS as an online survey before the first class and during the last week of class. Again, only the students who took the CLASS both before and after instruction were included in the data set (173 students total: 130 men, 43 women). The responses were analyzed using the template provided by the CLASS authors, which bundles the "strongly agree" and "agree" choices as the same answer, and bundles the "strongly disagree" and "disagree" choice as the same answer. Those answers are then compared to the "expert" response. The student's "percent favorable response," the percent of the student's responses that matched the expert response, is given as the score. The CLASS scoring scheme reports an overall score as well as scores in eight individual categories: personal interest, real world connection, problem solving general, problem solving confidence, problem solving sophistication, sensemaking or effort, conceptual understanding, and applied conceptual understanding. The shift is the change in a student's score over the course of the semester.

The effect size for the CLASS scores was calculated using Cohen's $d$ [Eq. (2)], where the mean overall percent favorable response was substituted in for the mean test score.

A paired $t$ test was used to determine the statistical significance of the difference in the pre and post FCI scores, as well as the difference in the pre and post CLASS scores. The standard error calculated throughout this paper is found by dividing the standard deviation of the sample by the square root of the number of samples. All protocols in the project were approved by the Virginia Tech Institutional Review Board (IRB- #16-171).

## IV. RESULTS

### A. Force Concept Inventory

In Table II, we present the average FCI pretest and post-test scores for both men and women, and the associated statistics. The average of the individual normalized gains for men and women on the FCI were 0.579 and 0.536, respectively, which is consistent with previous work documenting higher normalized learning gains in active-engagement classes than traditionally taught classes, where $\langle g \rangle < 0.3$ [3]. For both men and women, we found that the improvements were statistically significant ($p < 0.001$), and the effect size of the differences from pre- to post-scores, at 1.42 for men and 2.00 for women, is large ($d > 0.8$), according to Cohen [26].

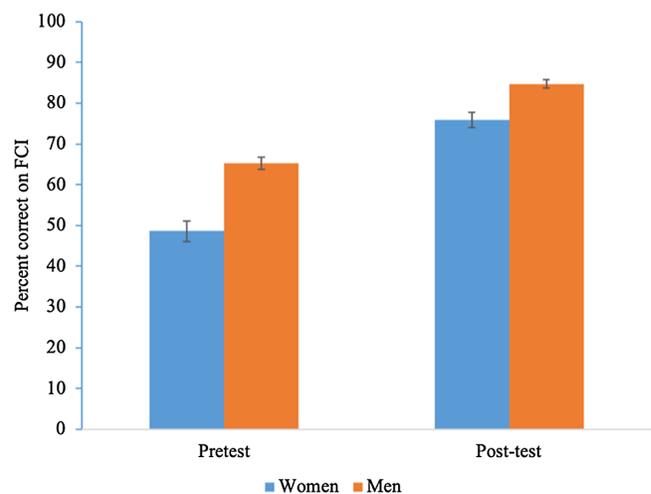

FIG. 1. Force Concept Inventory prescores and postscores (%) for men and women. Error bars represent the standard error of the mean.

TABLE III. Correlation of the normalized FCI gain and pretest FCI scores, including the regression coefficient $r$ and the $p$ value, for all students, men, and women.

|  | $r$ | $p$ value |
| --- | --- | --- |
| All students ($N = 217$) | 0.125 | 0.0670 |
| Men ($N = 168$) | 0.116 | 0.1346 |
| Women ($N = 49$) | 0.062 | 0.6741 |

Furthermore, our data showed a narrowing of the gender gap, the difference in average score for men and women, following instruction. Figure 1 shows that the gender gap prior to instruction was $16.7 \pm 3.0\%$, and after instruction, it dropped to $8.82 \pm 2.2\%$.

Previous studies have had mixed results in finding a correlation between FCI normalized gains and pretest scores. Although Hake [3] did not find a correlation, a larger study done by Coletta *et al.* [27], which included Hake's data, did find a positive correlation between FCI normalized gains and pretest scores. Using an $F$ test, we did not find a correlation (see Table III).

### B. Colorado Learning Attitudes about Science Survey (CLASS)

Figure 2 shows significant overall positive shifts in CLASS scores for men and women. The data in Table IV shows that the overall score for both genders shifted significantly ($p < 0.02$) to being more expertlike after instruction, a $+5.5$ shift for women and a $+3.9$ shift for men.

There were particularly strong positive shifts for both men and women in areas of problem solving and conceptual understanding. Table V displays the effect size of the overall shift, 0.406 for women and 0.337 for men. For reference, Cohen describes $d = 0.2$ to be a small effect and $d = 0.5$ to be a medium effect [26]. One anomaly in our data was the significant negative shift for men in the sense making or effort category. Although this is consistent with other findings [12], we are not sure why this category is an outlier in our data. This would be a possible topic of interest for future exploration.

### C. Correlations between FCI gain and CLASS scores

Although there has been much research on how instruction impacts student growth as measured by the FCI and the CLASS, the relationship between a student's score on attitudinal surveys, such as the CLASS or MPEX, and their normalized gain on conceptual inventories, such as the FCI or FMCE, is less well studied. Although some studies found a correlation between higher normalized gains on conceptual inventories with more expertlike scores on attitudinal surveys [28], it is usually small [29,30]. Moreover, the meta-analysis of students' beliefs about learning physics done by Madsen *et al.* [13] indicated that the relationship between students' conceptual learning gains and attitudinal shifts needs further study. To that end, we performed a linear regression $F$ test to find if there was a correlation between a student's normalized FCI gain and the shift in their CLASS overall and category scores (see Table VI). We then ran a similar analysis to find the correlation coefficients between a student's normalized FCI gain and their CLASS scores, both pre- and postinstruction, in Tables VII and VIII, respectively.

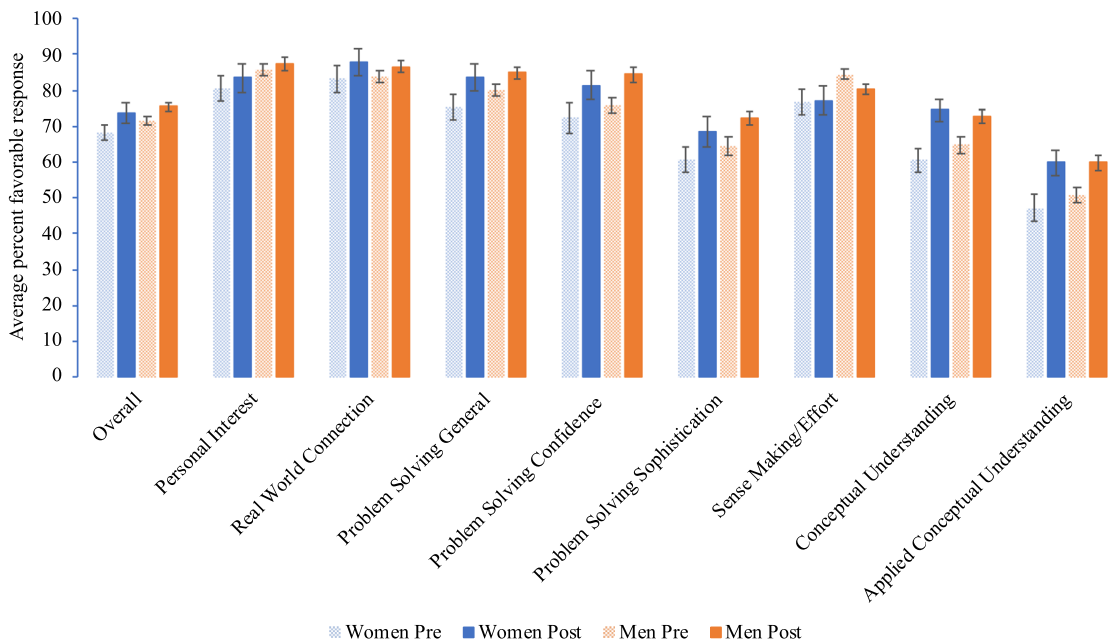

FIG. 2. CLASS pre- and postpercent favorable scores for men and women broken down by category. Error bars represent the standard error of the mean.

TABLE IV. Pre and post CLASS Scores (overall and category) for women and men. The shifts are measured by subtracting the preinstruction score from the postinstruction score. Scores are reported ± the standard error of the mean. The bolded values represent a large shift, as defined as larger than 2 standard errors.

| CLASS category | Women pre ($N = 43$) | Women post ($N = 43$) | Women shift | Men pre ($N = 130$) | Men post ($N = 130$) | Men shift |
|---|---|---|---|---|---|---|
| Overall | $68.2 \pm 2.1$ | $73.7 \pm 2.8$ | **$5.5 \pm 2.0$** | $71.6 \pm 1.1$ | $75.5 \pm 1.2$ | **$3.9 \pm 1.0$** |
| Personal interest | $80.6 \pm 3.6$ | $83.3 \pm 4.0$ | $2.7 \pm 3.8$ | $85.8 \pm 1.6$ | $87.3 \pm 1.7$ | $1.5 \pm 1.9$ |
| Real world connection | $83.1 \pm 3.7$ | $87.8 \pm 3.6$ | $4.7 \pm 4.5$ | $83.8 \pm 1.7$ | $86.6 \pm 1.7$ | $2.8 \pm 1.9$ |
| Problem solving general | $75.4 \pm 3.6$ | $83.4 \pm 3.7$ | **$8.1 \pm 3.9$** | $79.8 \pm 1.6$ | $84.7 \pm 1.5$ | **$4.9 \pm 1.4$** |
| Problem solving confidence | $72.3 \pm 4.2$ | $81.4 \pm 4.1$ | $9.1 \pm 4.7$ | $75.9 \pm 2.1$ | $84.4 \pm 2.0$ | **$8.5 \pm 2.1$** |
| Problem solving sophistication | $60.9 \pm 3.5$ | $68.6 \pm 4.3$ | $7.8 \pm 4.3$ | $64.5 \pm 2.4$ | $72.2 \pm 2.1$ | **$7.8 \pm 2.3$** |
| Sense making or effort | $76.9 \pm 3.6$ | $77.1 \pm 4.1$ | $0.2 \pm 3.5$ | $84.5 \pm 1.4$ | $80.2 \pm 1.6$ | **$-4.3 \pm 1.7$** |
| Conceptual understanding | $60.5 \pm 3.5$ | $74.4 \pm 3.2$ | **$14.0 \pm 4.0$** | $64.7 \pm 2.2$ | $72.5 \pm 1.9$ | **$7.8 \pm 2.2$** |
| Applied conceptual understanding | $47.2 \pm 3.8$ | $59.8 \pm 3.5$ | **$12.6 \pm 4.5$** | $50.6 \pm 2.2$ | $59.8 \pm 2.0$ | **$9.1 \pm 2.3$** |

For these analyses, we included only the students who had completed both the FCI and the CLASS before and after instruction, a total of 148 students: 110 men and 38 women. We considered a correlation to be both meaningful and statistically significant if $r > 0.3$ and $p < 0.05$ and bolded those correlations in the tables.

Interestingly, we found a large gender difference in how CLASS scores correlated with FCI gains: there were meaningful correlations for women, but we did not see any correlations for men. When looking at the women's CLASS shifts, only the sense making or effort category seemed to be correlated to the FCI gains ($r = 0.357$, $p = 0.028$). When analyzing the women's CLASS prescores, there were correlations with the FCI gains in three categories: personal interest ($r = 0.427$, $p = 0.0075$), problem solving general ($r = 0.336$, $p = 0.0389$), and problem solving confidence ($r = 0.332$, $p = 0.0418$), as well as the overall score ($r = 0.386$, $p = 0.0168$). We found correlations between the women's FCI gains and the CLASS postscores in almost every category: personal interest ($r = 0.413$, $p = 0.0099$), problem solving general ($r = 0.405$, $p = 0.0116$), problem solving confidence ($r = 0.342$, $p = 0.0353$), problem solving sophistication ($r = 0.348$, $p = 0.0325$) sense making or effort ($r = 0.457$, $p = 0.0039$), and conceptual understanding

TABLE V. Effect size and $p$ values for the post-pre shift for women and men in their overall percent favorable score. The 95% confidence interval (C.I.) on the effect is shown in the effect size.

| | Effect size | $p$ value |
|---|---|---|
| Women | 0.406 (−0.027, 0.840) | 0.0109 |
| Men | 0.337 (0.090, 0.584) | 0.00020 |

TABLE VI. Correlation of the normalized FCI gain and CLASS shifts, separated by category, for all students, men, and women. The regression coefficient $r$ and the $p$ value are given.

| | All students ($N = 148$) | | Men ($N = 110$) | | Women ($N = 38$) | |
|---|---|---|---|---|---|---|
| CLASS shift category | $r$ | $p$ value | $r$ | $p$ value | $r$ | $p$ value |
| Overall | 0.105 | 0.2035 | 0.084 | 0.3840 | 0.241 | 0.1450 |
| Personal interest | 0.076 | 0.3537 | 0.090 | 0.3487 | 0.020 | 0.9051 |
| Real world connection | 0.160 | 0.0516 | 0.177 | 0.0637 | 0.098 | 0.5593 |
| PS general | 0.054 | 0.5155 | 0.044 | 0.6503 | 0.109 | 0.5129 |
| PS confidence | 0.014 | 0.8643 | 0.013 | 0.8910 | 0.019 | 0.9091 |
| PS sophistication | 0.026 | 0.7521 | 0.009 | 0.9244 | 0.131 | 0.4339 |
| Sense making or effort | 0.161 | 0.0510 | 0.132 | 0.1695 | **0.357** | **0.0280** |
| Concept. under. | 0.002 | 0.9761 | 0.024 | 0.8049 | 0.156 | 0.3501 |
| App. concept. under. | 0.014 | 0.8688 | 0.029 | 0.7632 | 0.102 | 0.5423 |

TABLE VII. Correlation of the normalized FCI gain and CLASS prescores, separated by category, for all students, men, and women. The regression coefficient $r$ and the $p$ value are given.

| | All students ($N = 148$) | | Men ($N = 110$) | | Women ($N = 38$) | |
|---|---|---|---|---|---|---|
| CLASS prescore category | $r$ | $p$ value | $r$ | $p$ value | $r$ | $p$ value |
| Overall | 0.180 | 0.0295 | 0.406 | 0.1834 | **0.386** | **0.0168** |
| Personal interest | 0.087 | 0.2933 | 0.005 | 0.9558 | **0.427** | **0.0075** |
| Real world connection | 0.001 | 0.9902 | 0.033 | 0.7348 | 0.163 | 0.3284 |
| PS general | 0.157 | 0.0574 | 0.111 | 0.2490 | **0.336** | **0.0389** |
| PS confidence | 0.121 | 0.1424 | 0.071 | 0.4633 | **0.332** | **0.0418** |
| PS sophistication | 0.155 | 0.0592 | 0.137 | 0.1536 | 0.230 | 0.1649 |
| Sense making or effort | 0.114 | 0.1660 | 0.076 | 0.4299 | 0.228 | 0.1680 |
| Concept. under. | 0.157 | 0.0561 | 0.153 | 0.1108 | 0.151 | 0.3642 |
| App. concept. under. | 0.148 | 0.0732 | 0.147 | 0.1260 | 0.127 | 0.4482 |

TABLE VIII. Correlation of the normalized FCI gain and CLASS postscores, separated by category, for all students, men, and women. The regression coefficient $r$ and the $p$ value are given.

| | All students ($N = 148$) | | Men ($N = 110$) | | Women ($N = 38$) | |
|---|---|---|---|---|---|---|
| CLASS postscores category | $r$ | $p$ value | $r$ | $p$ value | $r$ | $p$ value |
| Overall | 0.235 | 0.0042 | 0.184 | 0.0551 | **0.449** | **0.0047** |
| Personal interest | 0.158 | 0.0550 | 0.094 | 0.3278 | **0.413** | **0.0099** |
| Real world connection | 0.176 | 0.0323 | 0.168 | 0.0802 | 0.239 | 0.1480 |
| PS general | 0.200 | 0.0149 | 0.150 | 0.1205 | **0.405** | **0.0116** |
| PS confidence | 0.137 | 0.0962 | 0.086 | 0.3712 | **0.342** | **0.0353** |
| PS sophistication | 0.192 | 0.0195 | 0.159 | 0.0962 | **0.348** | **0.0325** |
| Sense making or effort | 0.242 | 0.0031 | 0.190 | 0.0463 | **0.457** | **0.0039** |
| Concept. under. | 0.179 | 0.0291 | 0.151 | 0.1164 | **0.352** | **0.0302** |
| App. concept. under. | 0.149 | 0.0701 | 0.130 | 0.1774 | 0.268 | 0.1033 |

($r = 0.352$, $p = 0.0302$), as well as the overall score ($r = 0.449$, $p = 0.0047$).

### D. Correlations between FCI prescores and CLASS shifts

In addition to analyzing the correlations between FCI gains and CLASS scores, we also performed a linear regression $F$ test to see if there were any correlations between a student's FCI prescore and shift in their CLASS scores. We included the same students in this analysis as we did for the analysis of the correlations between a student's FCI gains and CLASS scores (Sec. IV. C), a total of 148 students: 110 men and 38 women. Again, we considered a correlation to be both meaningful and statistically significant if $r > 0.3$ and $p < 0.05$. Unlike the analyses in Sec. IV. C, however, we found no meaningful and statistically significant correlations between FCI prescore and CLASS shift (see Table IX).

TABLE IX. Correlation of the FCI prescore and CLASS shifts, separated by category, for all students, men, and women. The regression coefficient $r$ and the $p$ value are given.

| | All students ($N = 148$) | | Men ($N = 110$) | | Women ($N = 38$) | |
|---|---|---|---|---|---|---|
| CLASS shift category | $r$ | $p$ value | $r$ | $p$ value | $r$ | $p$ value |
| Overall | 0.087 | 0.2931 | 0.124 | 0.2007 | 0.155 | 0.3535 |
| Personal interest | 0.151 | 0.0666 | 0.191 | 0.0456 | 0.436 | 0.4093 |
| Real world connection | 0.198 | 0.0157 | 0.230 | 0.0158 | 0.109 | 0.5156 |
| PS general | 0.124 | 0.1338 | 0.090 | 0.3544 | 0.287 | 0.0812 |
| PS confidence | 0.031 | 0.7104 | 0.049 | 0.6106 | 0.031 | 0.7104 |
| PS sophistication | 0.018 | 0.8296 | 0.059 | 0.5384 | 0.254 | 0.1246 |
| Sense making or effort | 0.069 | 0.4036 | 0.171 | 0.0748 | 0.055 | 0.7421 |
| Concept. under. | 0.051 | 0.5420 | 0.036 | 0.7061 | 0.204 | 0.2195 |
| App. concept. under. | 0.082 | 0.3211 | 0.090 | 0.3486 | 0.180 | 0.2804 |

## V. DISCUSSION AND CONCLUSIONS

Although many interactive classes have shown strong learning gains for students, we attribute the atypical success of narrowing the gender gap on the FCI and obtaining positive attitudinal shifts for students of all genders at Virginia Tech to the confluence of interactive engagement pedagogies alongside a classroom culture that explicitly encourages collaboration, conceptual understanding, and a growth mindset in both the introductory physics and the seminar courses. It is important to note that our investigation was not a controlled study, and we are not suggesting that these methods will necessarily lead to the positive results that we have seen. Instead, we view our study as an existence proof that these positive results can occur in an introductory physics course for physics majors given the methods that we have outlined in this article.

We realize that gender gap analysis on the FCI is rife with complexities: the bias in the test itself [31], possible ceiling effects due to high post-test scores for men [32], as well as the underlying assumption that gender is binary and that the implicit goal is for women to perform like men [33]. Within our own research, future data collection would ask students to self-identify their gender identities. Nevertheless, we hope that our data can help illuminate practices that seem to be benefiting students of all genders in their study of physics.

Kreutzer and Boudreaux suggest that gender equity is promoted through the implementation of interactive engagement pedagogies in conjunction with instructional practices that attempt to foster domain (e.g., physics) belongingness [34]. Informed by Steele's work on "wise schooling" [35] and others [36–39], Kreutzer and Boudreaux outline instructional practices that support gender equity in the classroom:

- cultivate optimistic student-teacher relationships,
- affirm domain belongingness in women,
- practice nonjudgmental responsiveness,
- value multiple perspectives, and
- emphasize the expandability of knowledge.

Perhaps the classroom culture that the Virginia Tech instructors tried to create, one of fostering a growth mindset and a sense of belongingness, could be understood through Kreutzer and Boudreaux's framework. To encourage such a culture, the instructors

- learned all of their students' names;
- encouraged students to present alternative ideas, justifications, and solutions to conceptual and quantitative problems;
- explained the instructional value of student questions, intuition, mistakes, and incorrect answers;
- relayed stories of how former students who struggled with the material were able to master it after hard work and/or attending office hours, and reminded the current students of their own similar learning experiences;
- explained that the reason their class was using interactive engagement pedagogy was because they wanted the students to do well, and those strategies are based in physics education research;
- explained how the students would benefit from working collaboratively and articulating their thoughts to one another; and
- demonstrated that they wanted students to attend office hours by surveying the students about their availability and then scheduled office hours that minimized the students' conflicts.

Because much of the previous research has shown that courses with strong learning gains often do not have positive CLASS shifts, it is not surprising that we did not see a correlation between these two measures for the students as a whole. However, the correlations that we did see between the women's normalized FCI gains and their CLASS scores may indicate that the attitudes women have about learning physics may have a greater effect on their learning than the attitudes that men have on men's learning. Furthermore, we found that the women's CLASS postscore had a greater correlation to their FCI gains than their CLASS prescore. Apparently, what was most important in their conceptual growth was not the attitudes they arrived with, but rather, the attitudes that they formed over the semester.

Another result we found encouraging was that we saw no correlation between a student's FCI prescore and their CLASS shift. This result suggests that the positive CLASS shifts that we observed in the data is not strictly limited to students with a particular background in physics.

There may be multiple reasons why the students experienced positive attitudinal shifts, especially in the areas of conceptual understanding and applied conceptual understanding, in this study. Although the curricular focus on conceptual understanding was certainly one reason, we also think that the instructional practices of leveraging student intuition and reasoning [40,41] through responsive teaching [42] and guided mathematical sense making [43] were also essential. Perhaps all of these explicit discussions helped the students grow epistemologically—they developed a belief that they could reason through physics.

Our study demonstrates that an active engagement classroom with a blended pedagogical model of peer instruction, group problem solving, and direct instruction, along with an explicit focus on conceptual understanding and a growth mindset can result in high conceptual learning gains and positive attitudinal shifts for students of all genders. It is possible that our results are atypical because the students are physics majors in relatively small classes with other physics majors, but we are encouraged by the fact that students' attitudes toward learning physics are dynamic, and inching them forward may help reduce the gender gap. It is also possible that having a woman on the teaching team could have been a factor. Although our specific classroom conditions are unique, our findings may resonate with other researchers and practitioners who are investigating how to promote positive attitudinal shifts and gender equity in physics education.

## ACKNOWLEDGMENTS

The authors would like to thank Jill Sible for her thoughtful feedback and insights into this work and Jane Roberston-Evia for her advice on statistical analysis. This work was supported by a Math EAGER Grant (No. 1544225) from the National Science Foundation.